\pdfoutput=1
\documentclass[english,preprint,showpacs,preprintnumbers,amsmath,amssymb,aps,prb,superscriptaddress]{revtex4-1}

\newcommand{\ket}[1]{\ensuremath{\left|#1\right\rangle}}

\newcommand{\mathref}[1]{(\ref{#1})}
\newcommand{\figref}[1]{Fig.~\ref{fig:#1}}
\renewcommand{\vec}[1]{\ensuremath{\mathbf{#1}}}
\usepackage{hyperref}
\usepackage[usenames]{color}

\usepackage{graphicx,amsmath}

\begin{document}
\title{Magneto-absorption spectra of hydrogen-like yellow exciton series in cuprous oxide: excitons in strong magnetic fields}
\author{Sergey~L. Artyukhin}
\author{Dmitry Fishman}
\affiliation{Zernike Institute for Advanced Materials, University of Groningen, 9747 AG Groningen, The Netherlands.}
\author{Cl\'ement Faugeras}
\affiliation{Grenoble High Magnetic Field Laboratory, 25 Av. des Martyrs, 38 042 Grenoble, France}
\author{Marek Potemski}
\affiliation{Grenoble High Magnetic Field Laboratory, 25 Av. des Martyrs, 38 042 Grenoble, France}
\author{Alexandre Revcolevschi}
\affiliation{Institut de Chimie Mol\'{e}culaire et des Materiaux d'Orsay - UMR
8182, Universit\'{e} de Paris Sud, B\^{a}timent 410, 91405 Orsay Cedex, France}
\author{Maxim Mostovoy}
\author{Paul~H.M. van Loosdrecht}
\affiliation{Zernike Institute
for Advanced Materials, University of Groningen, 9747 AG Groningen,
The Netherlands.}

\date{\today}
\begin{abstract}
We study the absorption spectra of the yellow excitons in Cu$_2$O in high magnetic fields using polarization-resolved optical absorption measurements 
with a high frequency resolution.
We show that the symmetry of the yellow exciton results in unusual selection rules for the optical absorption of polarized light and that the 
mixing of ortho- and para- excitons in magnetic field is important. Our calculation of the energies of the yellow exciton series in an arbitrary 
magnetic field gives an excellent fit to experimental data and allows us to understand the complex structure of excitonic levels and their 
magnetic field dependence, which resolves the old-standing disagreement between the results of optical absorption and cyclotron resonance measurements.
\end{abstract}

\maketitle
\hyphenation{mag-ne-to-ab-sorp-tion}
\section{Introduction}
Cuprous oxide Cu$_2$O was the first material in which Wannier-Mott excitons \cite{Wannier37} -- the particle-hole pairs bound by the Coulomb 
interaction -- were observed. These states give rise to hydrogen-like series of absorption lines in the optical absorption spectrum of Cu$_2$O at 
the photon energies described by the Rydberg formula, 
${E_n=E_\text{gap}-{\rm Ry_X}/n^2}$, where ${\rm Ry_X}=98$ meV is the excitonic Rydberg constant and $E_\text{gap}=2.17$ eV is the optical gap. 
The only exception is the 
$n = 1$ exciton, which is barely visible because the valence and conduction bands of this material have the same parity \cite{Elliott61}. Due to the small 
size of the $n = 1$ exciton, its energy 
is strongly affected by the exchange and central cell corrections \cite{Kavoulakis97}.
The revival of interest in Cu$_2$O was motivated by the search for the Bose-Einstein condensation of the exciton gas 
\cite{Keldysh68,Kuwata11,Snoke02}. These studies underscored the importance of a quantative description of excitons in this material.
Despite the recent progress \cite{Kavoulakis97,Snoke95}, a number of fundamental problems remain, surprisingly, unsolved.  For example, there is 
a siginificant discrepancy between the effective masses of electrons and holes deduced from the optical meausrements \cite{Halpern67, zhil69} and 
the cyclotron resonance experiments \cite{hodby76}. In addition, a detailed interpretation of the magnetooptical spectra is still lacking 
\cite{sas73,kob89,sey03}. In this paper we address these issues and resolve the descripancy using the high-resolution measurements of the low-energy 
magnetooptical absorption spectra of Cu$_2$O combined with the numerical calculations of these spectra in the regime of intermediate magnetic fields.

\begin{table}[b]
\begin{center}
\begin{tabular}{|l|l|l|l|l|l|l|}
\hline
method			&$m_e$	&$m_h$	 	&$\mu$&$\epsilon$&$g_c$&$g_v$\\
\hline
optical\cite{zhil69}	&0.61	&0.84			&0.35&7.1&2.0&0.28\\
cyclotron resonance \cite{hodby76}&0.99	&0.69&0.41&7.5&&\\
\hline
\end{tabular}
\caption{\label{tab:masses}Effective masses of electrons $m_e$, holes $m_h$, the dielectric constant $\epsilon$ and an exciton reduced mass $\mu$ 
obtained in different experiments. The masses are in the units of bare electron mass $m_0$. The last two columns give the $g$-factor of the electrons 
and holes, respectively.}
\end{center}
\end{table}
Optical absorption spectra measured in zero fields give information about the excitonic Rydberg constant and the reduced mass of the electron-hole pair. 
Further information can be obtained from the splittings of excitonic levels in applied  electric and magnetic fields. The magnetoabsorption spectra of 
excitons in Cu$_2$O were extensively studied over the past decades \cite{gross56, Halpern67, sey03, kob89, ham00, sas73}.

Because of the large dielectric constant and small effective masses of charge carriers in Cu$_2$O, the exciton radius greatly exceeds that of a hydrogen 
atom, which amplifies the effects of external electric and magnetic fields on the exciton wave function. For example, the exciton ionization in Cu$_2$O  
occurs in an electric field of $E=5$ kV/cm \cite{gross56}, whereas the characteristic field required to ionize the hydrogen atom is of the order of 
$E\sim 1000$ kV/cm. Pronounced and fairly complex Zeeman and Stark effects were first observed by Gross and Zakharchenya in magnetic fields up to 2.8 T \cite{gross56}.

Due to the large radius of the Cu$_2$O excitons, the regime of strong magnetic fields, in which the field-induced level splittings become 
comparable with the zero field level spacing, can be easily reached in a laboratory (for hydrogen atom such strong fields are only found in 
neutron stars). For a magnetic field of $H=30$~T the cyclotron energy is comparable to the binding energy already for the $n=3$ exciton. 
For higher levels, the strong-field regime is reached at even lower fields ($\beta_n = \hbar\omega_c/(2Ry_X/n^2)\approx 0.05 n^2$, where $omega_c$ is 
the cyclotron frequency). However, the interpretation of experimentally measured spectra of exciton states with large $n$ in applied magnetic fields 
is complicated by the large number of overlapping lines, which makes it difficult to extract exciton parameters from such spectra in a reliable way. 

Zhilich $et~al.$ \cite{zhil69} studied the oscillations of optical absorption well above the gap in magnetic field up to $10$~T. These oscillations 
originate from the transitions between the Landau levels of electrons and holes. The effective masses of electrons and holes were estimated to be 
$m_e=0.61m_0, m_h=0.84m_0$, where $m_0$ is the bare electron mass. The accuracy of these results was limited by poor energy resolution and 
available magnetic fields. 

At lower energies, in the region of bound excitons, the absorption lines are much sharper and better suitable for extraction of exciton parameters. 
Sasaki and Kuwabara \cite{sas73} measured the magnetoabsorption spectrum in static magnetic fields up to $16$~T. Kobayashi $et~al.$ \cite{kob89} 
studied the $n=2$ and $n=3$ exciton absorption in pulsed magnetic fields up to $150$~T. Seyama {\it et al.} \cite{sey03} measured the 
spectra in static fields up to $25$~T with better spectral resolution. However, the complexity of the spectra with large number of overlapping lines 
prevented the unambiguous assignment of excitonic levels. In addition, Coulomb interactions between the electron and hole were not taken into account 
in the analysis of the spectra.

The effective masses obtained from cyclotron resonance experiments \cite{hodby76}:  
$0.58m_0$ and $0.69m_0$ for light and heavy holes, respectively, and $0.99m_0$ for electrons are significantly different from the masses obtained in 
magnetooptical measurements (see Table \ref{tab:masses}). This disagreement was ascribed to polaronic effects \cite{Halpern67, zhil69}.
More recently the value of $0.575m_0$ was derived 
from pulsed cyclotron resonance experiments\cite{Naka12}.

In the attempts to extract 
the exciton parameters from the spectra measured at high magnetic fields, the Coulomb energy of electron and hole was 
neglected; an approximation which is justified only for very large magnetic fields and large-$n$ excitons. In works using 
low excitonic levels the magnetic field was treated perturbatively; an approximation only justified in a limited area of the spectra.

Therefore, the most promising are the levels with $n\leq 5$, falling into intermediate field regime $\beta\sim 1$. There is no small parameter for a 
perturbative expansion, and numerical calculations are required to obtain the exciton spectrum. 

We performed polarized high spectral resolution optical absorption measurements in static magnetic field up to $32$~T and obtained with high accuracy 
the magnetic field dependence of the exciton energies for $n = 2,3$ and $4$. The largest part of the spectrum lies in the intermediate-field region, 
in which the interaction of electrons and holes with the magnetic field is comparable with Coulomb interaction, so that neither of these interactions 
can be treated perturbatively. We calculated the exciton energies in the intermediate regime numerically and extracted the effective masses and 
$g$-factors of the electron and hole by fitting the data.  Not only do we get a good agreement between the theory and magnetoabsorption experiments, 
but the masses that we obtain coincide with those obtained from the cyclotron resonance experiments, thus resolving the long-standing contradiction. 

This paper is organized as follows.
In Sec. \ref{sec:exp} we present the results of magneto-absorption measurements in Cu$_2$O in static magnetic fields up to $32$~T.
In Sec. \ref{sec:symmetry} we discuss crystal symmetry and the band structure of Cu$_2$O. We determine the symmetry of electron and hole wave 
functions, which allows us to derive the optical selection rules (see Sec. \ref{sec:selection}). 
Then, in Sec. \ref{sec:relative} we calculate the peak positions in the absorption spectrum. The comparison between the experimental and 
theoretical results is discussed in Sec.  \ref{sec:discussion}.

\section{\label{sec:exp}Experimental results}
The magneto-absorption of Cu$_2$O has been studied in a Faraday geometry ($\textbf{H}||\textbf{k}$) with magnetic fields up to $B=32$~T at a 
temperature of $T=1.2$~K. For the experiments, platelets (thickness 40~$\mu$m) cut and polished from floating zone grown Cu$_2$O single 
crystals \cite{sch74} of [100] orientation were placed in a pumped liquid helium bath cryostat. A halogen lamp was used as a light source. 
Circular polarization was achieved by a combination of a quarter-wavelength plate and a polarizer situated inside the cryostat. 
The polarized light then was detected with a double monochromator (resolution 0.02 nm) equipped with a LN$_2$ cooled CCD camera. Right ($\sigma^+$) 
and left ($\sigma^-$) circular polarization of transmitted light was resolved by switching the magnetic field direction: 
($\textbf{H}\uparrow\uparrow\textbf{k}$) for $\sigma^+$ polarization detection and ($\textbf{H}\uparrow\downarrow\textbf{k}$) for $\sigma^-$ 
polarization detection.

\begin{figure}
\begin{center}
\includegraphics[width=8cm]{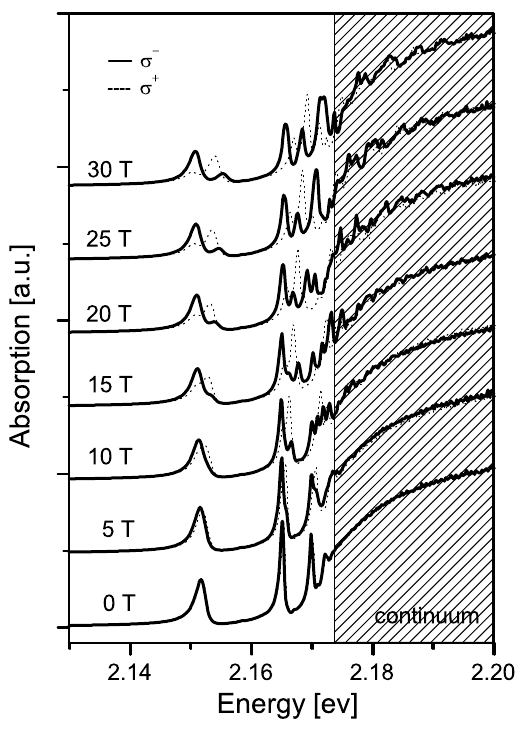}
\end{center}
\caption{\label{fig1}Magneto-absorption spectra of Cu$_2$O in yellow exciton energy range for different magnetic field strengths. 
Solid line - left circular polarized spectra; dotted line - right circular polarized spectra. Bath temperature $T=1.2$~K. 
The spectra for different fields have been given an offset for clarity.}
\end{figure}

\begin{figure*}[ht]
\begin{center}
\includegraphics[width=16cm]{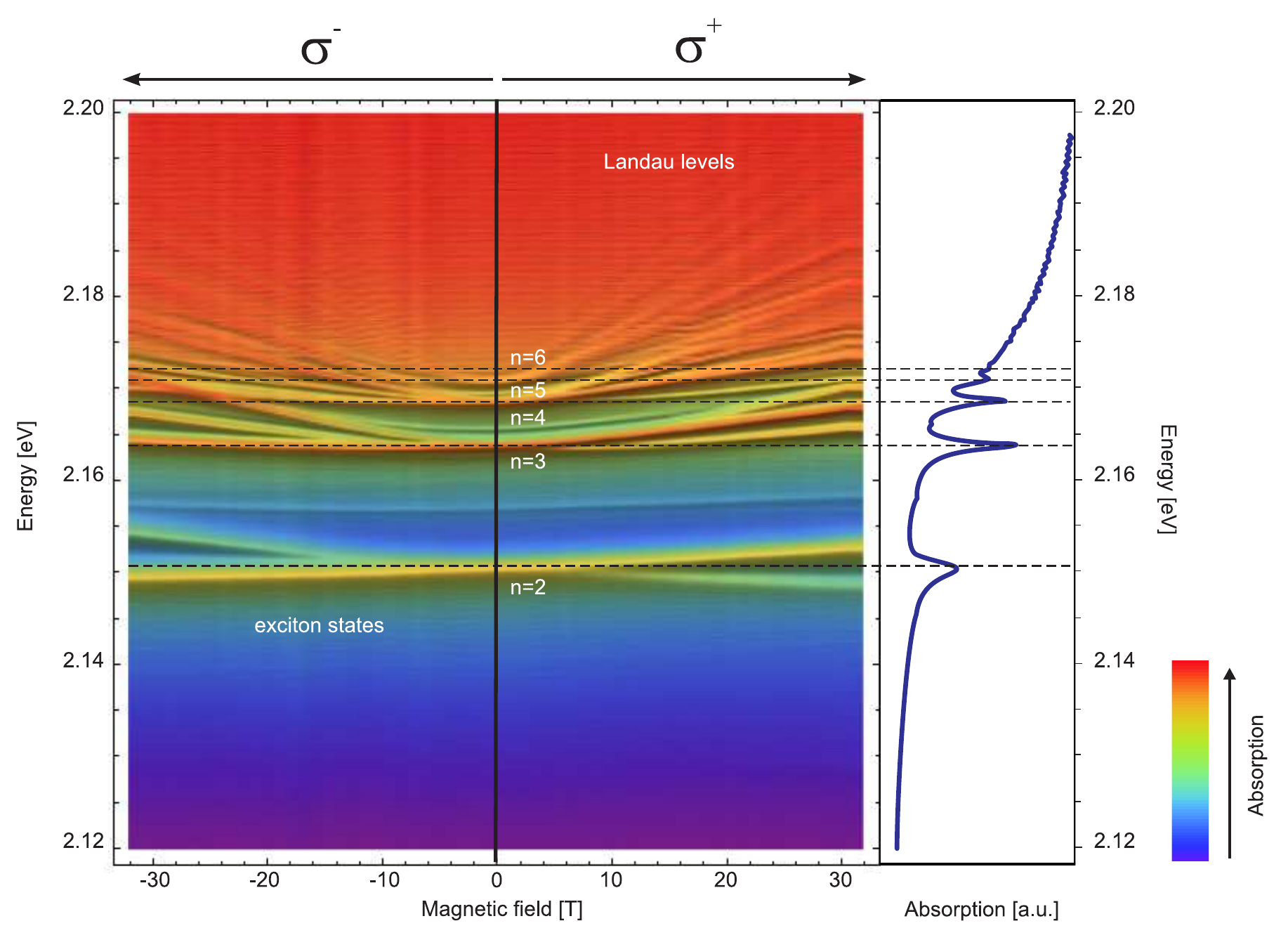}
\end{center}
\caption{\label{fig2} (color online) Image plot of the absorption spectra of Cu$_2$O at $T=1.2$~K.
Left part - $\sigma^-$ polarization; right part - $\sigma^+$
polarization. The red parts are showing the higher absorption
areas.}
\end{figure*}

Figure~\ref{fig1} shows the magnetic field dependence of the absorption spectra for some selected field strengths. In the absence of a 
magnetic field, the absorption spectrum of Cu$_2$O exhibits the well known hydrogen-like absorption series below the inter band transition energy. 
The spectral resolution of the experiments and the quality of the sample allows the observation of at least 5 exciton peaks of the yellow series ($n$=2-6). 
Upon applying a magnetic field, the spectra become increasingly more complex; The exciton absorption peaks show a progressive splitting and the continuum 
above the band gap energy shows a complex magneto-oscillatory spectrum originating from Landau quantization of the unbound electron and hole states.

In order to address the complexity of the spectra, a detailed field dependence of the circular polarized spectra up to B=32 T was performed. 
Figure~\ref{fig2} represents an overview of the optical absorption 
spectra in a false color representation of the intensity as a function of the photon energy and magnetic field. 
The left part represents $\sigma^-$ 
spectra, whereas the right part represents $\sigma^+$ spectra.
For the $n=2$ state, the absorption peak shows a splitting continuous combined 
Zeeman and Langevin shift upon increasing magnetic field throughout the
whole magnetic field range. 

The absorption spectrum becomes more complex with the increase of the principal quantum number $n$. In case of Cu$_2$O, only transitions to 
p-states for each principal quantum number $n$ are dipole-allowed. These states are clearly observed in the absence of a magnetic field. With increasing 
magnetic field, one can observe other $l$-index (orbital number) states, due to the finite off-diagonal elements induced by the magnetic field, \cite{kob89}. 
The general behavior of the absorption peaks in this area are described in \cite{sey03}. However, the diamagnetic coefficients of these peaks cannot 
be explained using the simple calculation based on first-order perturbation theory\cite{sey03}. For $n=3$ the situation is still relatively simple 
in that only three lines are observed which do not show any additional splitting upon increasing field strength. For larger $n$ the exciton peaks 
cross or show avoided crossing behavior leading to deviations of the expected diamagnetic and Zeeman shifts\cite{Halpern67, sey03, kob89, sas73}. 
Furthermore, additional absorption lines appear at high magnetic fields which will be discussed later.

The energy region in the vicinity and above the band-gap energy is of particular interest (Fig.~\ref{fig2}): already at 8T equidistant 
quasi-Landau levels become visible. Hammura $et~al.$ \cite{ham00} suggested, that the electron-hole pair undergoes a periodic orbit mainly 
determined by the Coulomb potential which is perturbed by the presence of a magnetic potential. Seyama $et.~al.$ \cite{sey03} considered 
these levels as a result of frequent level crossing of states with different quantum numbers $n$ and $l$, since the level spacing is 
comparable with the anti-crossing gaps. As described in 
the remainder of this paper, the proper description of the complex magnetoabsorption spectrum Fig.~\ref{fig2} follows directly from the gradual 
transition from exciton to the magneto-exciton behavior without the need for the orbital interference effects as described in \cite{ham00}.

\section{\label{sec:symmetry}Symmetry of the yellow excitons}

\begin{figure}[ht]
 \centering
\includegraphics[width=.8\columnwidth]{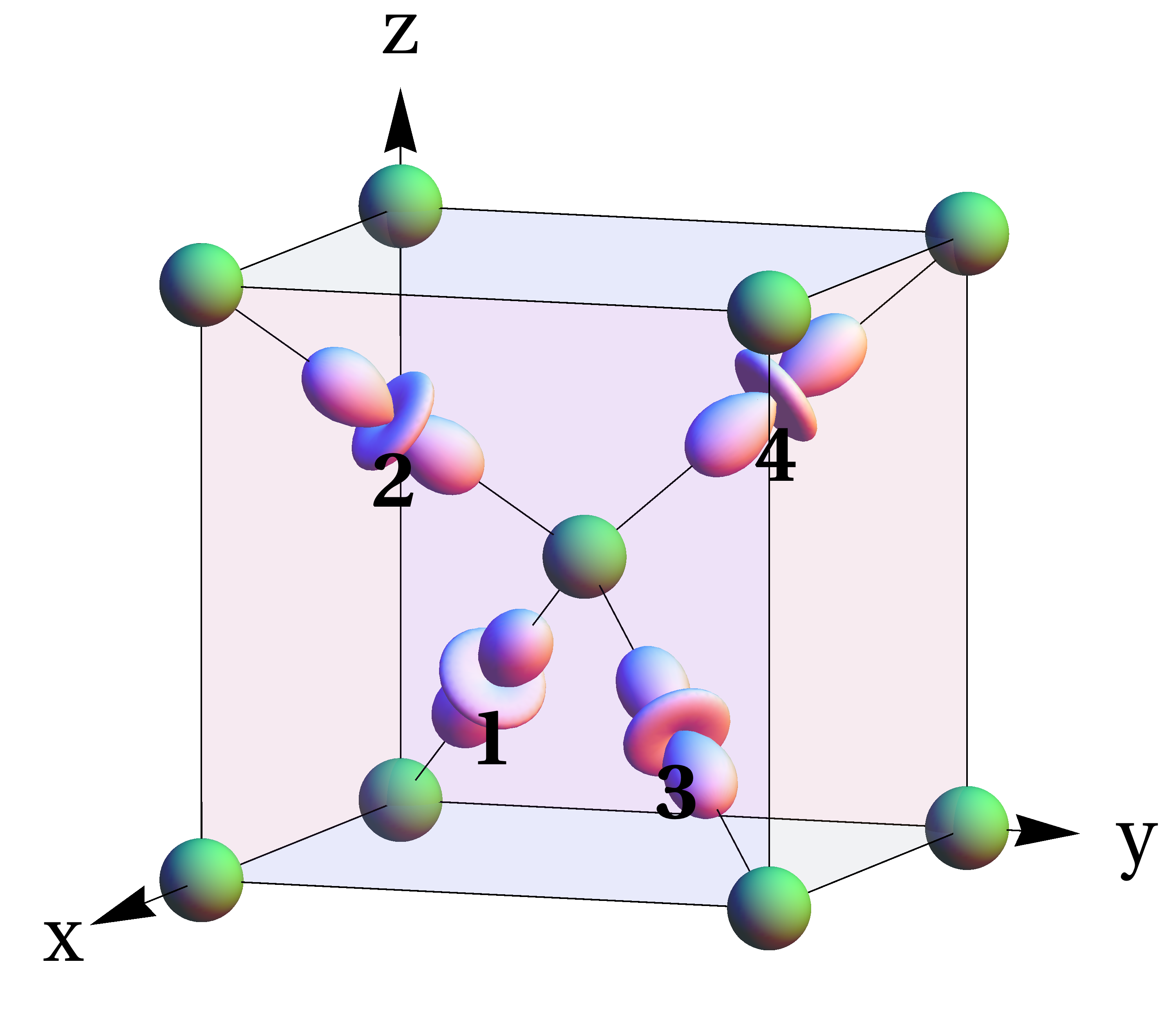}
 \caption{(color online) The unit cell of Cu$_2$O. Oxygen ions in green form
a body-centered cubic lattice. The Cu $d$ orbitals (numbered) give the dominant contribution to the upper valence band.}
 \label{fig:UnitCell}
\end{figure}

The cuprite $\rm Cu_2O$ has a cubic symmetry (space group \textit{Pn${\bar 3}$m}) with 4 Cu ions 
in the unit cell (see \figref{UnitCell}). The electron can be excited from the highest valence band, 
formed mostly by the Cu $3d$ orbitals, to the lowest conduction band, formed by the Cu $4s$ orbitals. 
The yellow excitons are then formed by binding the excited electrons and holes with Coulomb interaction.

The symmetries of the electron and hole bands, respectively, $\Gamma^{+}_6$ and $\Gamma_7^{+}$ \cite{Elliott61}, 
can be understood as follows. Each Cu$^{+}$ ion is coordinated by two oxygen ions in the dumbbell configuration, which splits its $d$-shell 
into two doublets,$(x_i^2-y_i^2,x_iy_i)$ and $(x_iz_i,y_iz_i)$, and one singlet, $3z_i^2-r_i^2$. Here the direction of the $z_i$ axis is 
parallel to the O-Cu-O line passing through the $i$-th Cu ion ($i=1,2,3,4$) in the unit cell and is different for different Cu sites 
(see \figref{UnitCell}). Since the $3z_i^2-r_i^2$ state has the highest energy, we assume for simplicity that the upper valence band is 
formed by these orbitals only.

As the hopping amplitudes between all pairs of the $3z_i^2-r_i^2$ orbitals on neighboring Cu sites are equal by symmetry, the tight-binding 
band structure at the $\Gamma$-point consists of the non-degenerate singlet state,
\begin{equation}
\ket{S} = \frac{1}{2} \left(\vert 1\rangle + \vert 2\rangle + \vert 3\rangle + \vert 4\rangle \right)
\label{eq:singlet}
\end{equation}
and the triplet of degenerate states,
\begin{equation}
\left\{
\begin{array}{rcl}
\vert X \rangle &=& \frac{1}{2} \left(+\vert 1\rangle - \vert 2\rangle - \vert 3\rangle + \vert 4\rangle \right),\\ \\
\vert Y \rangle &=& \frac{1}{2} \left(-\vert 1\rangle + \vert 2\rangle - \vert 3\rangle + \vert 4\rangle \right),\\ \\
\vert Z \rangle &=& \frac{1}{2} \left(-\vert 1\rangle - \vert 2\rangle + \vert 3\rangle + \vert 4\rangle \right),
\end{array}
\right.
\label{eq:triplet}
\end{equation}
where $\vert i\rangle$ denotes the $3z_i^2-r_i^2$ orbital on the $i$-th Cu site. Table~\ref{tab:t2g} shows the transformation of these 
three states under the generators of $Pn{\bar 3}m$ group: the $\pi$-rotation around the $z$-axis, $C_{2z}$: $(x,y,z) \rightarrow 
\left(\frac{1}{2}-x,\frac{1}{2}-y,z\right)$, the $\frac{2\pi}{3}$-rotation around the body diagonal of the cube, $C_{3}$: $(x,y,z) 
\rightarrow (z,x,y)$,  the mirror,  $m_{x-y}$: $(x,y,z) \rightarrow (y,x,z)$, and inversion $I$: $(x,y,z)\to(1/2-x,1/2-y,1/2-z)$.

\begin{table}
\begin{center}
\begin{tabular}{|c|c|c|c|c|}
\hline
State & $C_{2z}$ & $C_3$ & $m_{x-y}$ & $I$ \\
\hline
$\vert X \rangle$ & $- \vert X \rangle$& $\vert Y \rangle$ & $\vert Y \rangle$ & $\vert X \rangle$ \\
$\vert Y \rangle$ & $- \vert Y \rangle$ & $\vert Z \rangle$ & $\vert X \rangle$ & $\vert Y \rangle$  \\
$\vert Z \rangle$ & $+ \vert Z \rangle$ & $\vert X \rangle$ & $\vert Z \rangle$ &  $\vert Z \rangle$ \\
\hline
\end{tabular}
\caption{\label{tab:t2g}The transformation rules of the $\ket{X},\ket{Y},\ket{Z}$ states under the operations of $Pn{\bar 3}m$ group.}
\end{center}
\end{table}

As the hopping between nearest-neighbor Cu sites is mediated by oxygen ions, the hopping parameter $t$ of the effective tight-binding model 
describing the Cu sites only, $$H = -t \sum_{<i,j>\sigma }\left(c^\dag_{i\sigma}c_{j \sigma} + c^\dag_{j \sigma}c_{i\sigma}\right),$$ where 
the operator $c_{i \sigma}$ annihilates electron in the state $3z_i^2 - r_i^2$ on the site $i$ with the spin projection $\sigma$, is given 
by $t = \frac{t_{pd}^2}{\Delta}$, where $t_{pd}$ is the hopping amplitude between the Cu to O sites and $\Delta>0$ is the charge transfer 
energy. Since $t > 0$, the states with the energy $+2t$ at the $\Gamma$-point lie higher than the singlet state with the energy $-6t$. 
The spin-orbit interaction further splits the six (including the spin degeneracy) states into a doublet \cite{Kavoulakis97},
\begin{equation}\label{eq:valence}
\left\{
\begin{array}{rcl}
\ket{\uparrow}_v&=&-\frac{1}{\sqrt{3}}\left[(\ket{X}+i\ket{Y})\ket{\downarrow}+
\ket{Z}\ket{\uparrow}\right],\\ \\
\ket{\downarrow}_v&=&\frac{1}{\sqrt{3}}\left[(-\ket{X}+i\ket{Y})\ket{\uparrow}+\ket{Z}\ket{\downarrow}\right],
\end{array}
\right.
\end{equation}
and a quadruplet (here the subscript $v$ indicates the valence band). The energy of the doublet is higher than the energy of the quadruplet by $\sim 134$
meV \cite{Uihlein81}. This spin-orbital splitting  originates from the virtual admixture of $x_iz_i$ and $y_iz_i$ states to the $3z_i^2 - r_i^2$ 
state by the spin-orbit coupling on Cu sites. The doublet belongs to the upper valence band, which gives rise to the yellow exciton, while the 
quadruplet gives rise to the green exciton series \cite{gross56}. We stress that our $\ket{X}$, $\ket{Y}$ and $\ket{Z}$ states are formed 
by the $3z_i^2 -r_i^2$ orbitals of 4 Cu ions in the unit cell and are different from the atomic $xy$, $yz$, and $zx$ orbitals discussed 
in Ref.~\onlinecite{Kavoulakis97}. 

Using Eq.~(\ref{eq:valence}) and Table~\ref{tab:t2g}, one finds that the valence-band doublet,
$
\psi_v = \left(
\begin{array}{c}
\ket{\uparrow}_{v}\\
\ket{\downarrow}_{v}
\end{array}
\right)
$, transforms as a $\Gamma^{+}_7$ representation:
\begin{equation}
\begin{array}{l}
C_{2z} \psi_{v} =
e^{-i\frac{\pi}{2}\sigma_z} \psi_{v} =
\left(
\begin{array}{cc}
-i & 0\\
0 & i
\end{array}
\right)
\psi_{v},\\
C_{3} \psi_{v} = e^{-i \frac{\pi}{3\sqrt{3}}
\left(\sigma_{x} + \sigma_{y} + \sigma_{z}\right)}
\psi_{v} =
\frac{1}{2}
\left(
\begin{array}{cc}
1-i & -1-i\\ \\
1-i & 1+i
\end{array}
\right)
\psi_{v},\\ \\
m_{x-y} \psi_{v} =
\frac{i}{\sqrt{2}}
\left(\sigma_{x}-\sigma_{y}\right)
\psi_{v} =
\frac{i}{\sqrt{2}}
\left(
\begin{array}{cc}
0 & 1+i\\ \\
1-i & 0
\end{array}
\right)
\psi_v,\\ \\
I \psi_{v} = \psi_{v}.
\end{array}
\label{eq:transvalence}
\end{equation}

Similarly, the lowest conduction band, formed by the Cu $4s$ orbitals, splits into a triplet and singlet at the $\Gamma$-point with the 
singlet state having a lower energy. Since the orbital part of the singlet wave function [see Eq.~(\ref{eq:singlet})], is invariant 
under all operations of the space group, the symmetry of the doublet, $\psi_{c} = \left(
\begin{array}{c}
\ket{\uparrow}_{c}\\
\ket{\downarrow}_{c}
\end{array}
\right)$, formed by the spin-up and spin-down electron states in the lowest conduction band, is determined by its spin wave function. 
Thus, the conduction electron in the yellow exciton has the same transformation  properties as the valence electron 
[see Eq.~(\ref{eq:transvalence})], except for the opposite sign for the mirror transformation, $m_{x-y}$, and, hence, belongs to 
$\Gamma^{+}_6$ representation.

Finally, the conduction electron and the valence hole form
ortho- and para-excitons with the total spin, $S$, respectively, 1 and 0. Due to the exchange interaction between the 
conduction and valence electrons in the $n = 1$ yellow exciton state, the energy of the ortho-exciton is $12$meV higher 
than that of para-exciton \cite{kiselev71,pikus71,denisov73,Kavoulakis97,Fishman09}.

\section{\label{sec:selection}Selection rules}

Since the valence $3d$ and conduction $4s$ bands have the same parity, the excitation of the yellow exciton series is dipole forbidden 
and results from the electric quadrupole transition \cite{Elliott61}. The conduction and valence band doublets, $\psi_c$ and $\psi_v$, 
transform under the mirror $m_{x-y}$ with opposite signs (see Sec.~\ref{sec:symmetry}), resulting in the``wrong" symmetry of yellow 
excitons: the paraexciton wave function, 
\[
\ket{S=0,S_z=0} = \frac{1}{\sqrt{2}}\left(\psi_{c\uparrow}\psi_{v\downarrow} -\psi_{c\downarrow}\psi_{v\uparrow} \right),
\]
is odd under $m_{x-y}$, while the orthoexciton wave function with zero projection of the total spin, 
\[
\ket{S=1,S_z=0} = \frac{1}{\sqrt{2}}\left(\psi_{c\uparrow}\psi_{v\downarrow} +\psi_{c\downarrow}\psi_{v\uparrow} \right),
\]
is even. 

The invariance of the paraexciton wave function $\ket{0,0}$ under $C_3$ and $C_2$ rotations requires that the amplitude of the 
photoexcitation of this state has the form,
\begin{equation}
A_{00} \propto \sum_{\vec{k}} \varphi_{\vec{k}}^{\ast}\left(\vec{e} \cdot \vec{k}\right),
\end{equation}
where $\vec{k}$ is the relative wave vector of the electron-hole 
pair, $\varphi_{\vec{k}}$ is the wave function of the relative 
motion, discussed in the next section, and $\vec{e} = \vec{e}_{\vec{q}\lambda}$ 
is the polarization vector of the photon with the wave vector ${\bf q}$ 
and polarization $\lambda$. The scalar product $\left(\vec{e} \cdot \vec{k}\right)$ 
is invariant under $m_{x-y}$, while $A_{00}$ must be odd, implying that $A_{00} = 0$, i.e., paraexcitons cannot be excited via 
the one-photon absorption.

The orthoexciton states $\ket{1,S_z}$ with $S_z = -1,0,1$, are excited by the components of the quadrupolar tensor,
\begin{equation}
Q_{ab} \propto \sum_{\vec{k}} \varphi_{\vec{k}}^{\ast}\left(e_a k_b + e_b k_a - \frac{2}{3}\delta_{ab} \vec{e} \cdot \vec{k} \right).
\end{equation}
The excitation amplitudes, invariant under all crystal symmetries, have an obvious form for the Cartesian components of the 
orthoexciton atomic wave functions, $\ket{{\rm x}}$, $\ket{{\rm y}}$, and $\ket{{\rm z}}$:
\begin{equation}
\left\{
\begin{array}{lcc}
\ket{1,1} & = & - \frac{1}{\sqrt{2}}\left(\ket{{\rm x}}+i\ket{{\rm y}}\right),\\ \\
\ket{1,0} & = & \ket{{\rm z}},\\ \\
\ket{1,1} & = & \frac{1}{\sqrt{2}}\left(\ket{{\rm x}}-i\ket{{\rm y}}\right).
\end{array}
\right.
\end{equation}
The form of the invariant amplitudes is: 
\begin{equation}
\begin{array}{lcc}
A_{\rm x} & \propto & Q_{yz},\\
A_{\rm y} & \propto & Q_{zx},\\
A_{\rm z} & \propto & Q_{xy},
\end{array}
\end{equation}
and the proportionality coefficient is the same for all states.
 
For the Faraday geometry, $q \| H$ (and  $H \| z$), 
\begin{equation}
A_x,A_y \propto \sum_{\vec{k}} k_z \varphi^{\ast}_{\vec{k}} \propto \left.\frac{\partial \varphi^{\ast}}{\partial z}\right|_{\vec{r} = 0},
\label{eq:selrule1}
\end{equation}
so that these amplitudes are only nonzero for $m = 0$, where $m$ is the $z$-projection of orbital momentum of the relative motion of the 
electron-hole pair. Similarly, $A_z$, does not vanish only for $m = \pm 1$ states with nonzero $\left[
\frac{\partial \varphi^{\ast}}{\partial x} 
\mp i 
\frac{\partial \varphi^{\ast}}{\partial y}\right]_{\vec{r} = 0}$. 
For zero magnetic field the allowed excited states have the orbital momentum $l = 1$ ($p$-states).   

In this way we can obtain the following unusual selection rules for orthoexcitons from the yellow series: a photon with the 
polarization $\lambda = \pm 1$ [$\vec{e}_{\vec{q},\pm 1} = \frac{1}{\sqrt{2}}\left(1,\pm i,0\right)$] excites either the state 
with $S_z  = -\lambda$ and $m = 0$, or the state with $S_z = 0$ and $m = - \lambda$. These selection rules are opposite to those 
for rotationally-invariant systems, where the $z$-component of the total angular momentum is a good quantum number.

\section{\label{sec:relative}Motion of electron-hole pair in magnetic field}

The atomic part of the exciton wave function, discussed in the previous section, remains largely unaffected by an applied 
magnetic field of $32$T, except for the mixing of the paraexciton and orthoexciton states. On the other hand, magnetic field 
has a strong effect on the relative motion of the electron and hole, especially, in highly-excited excitonic states. The problem of 
finding energies of excitonic states in magnetic field is simplified by the conservation of the total momentum of the electron-hole 
pair \cite{GorDzya1968}, which makes it equivalent to the problem of hydrogen atom in magnetic field \cite{lozovik03}, 
\footnote{for a review see I.B. Khriplovich, G.Yu. Ruban, arXiv:quant-ph/0309014v2}.

The relatively slow motion of electron and hole in the Cu$_2$O excitonic states is, to a good approximation, decoupled 
from the dynamics of their spins and can be considered separately. The Lagrangian describing this  motion is
\begin{eqnarray}
&~&L =\frac{m_e\dot{\vec{r}}_e^2}{2}
+\frac{m_h\dot{\vec{r}}_h^2}{2}
-\frac{e}{c}\vec{A}(\vec{r}_e)\cdot\dot{\vec{r}}_e
+\frac{e}{c}\vec{A}(\vec{r}_h)\cdot\dot{\vec{r}}_h
\nonumber
\\&~& +\frac{e^2}{\varepsilon \left|\vec{r}_e - \vec{r}_h\right|},
\end{eqnarray}
where $\vec{r}_e$($\vec{r}_h$) is the electron(hole) coordinate, $m_e$ and $m_h$ are the electron and hole masses, and 
$\vec{A}(\vec{r})=\frac{1}{2}[\vec{H}\times \vec{r}]$ is the vector potential (the electron charge is $-e$).

In the center-of-mass and relative coordinates,
$\vec{R}=\frac{\vec{r}_em_e+\vec{r}_hm_h}{m_e+m_h}$ and $\vec{r}=\vec{r}_e-\vec{r}_h$, the Lagrangian has the form,
\begin{equation}
L=\frac{M\dot{\vec{R}}^2}{2} +  \frac{\mu\dot{\vec{r}}^2}{2} - \frac{e}{c}\left(\dot{\vec{R}} + \frac{\gamma}{2} \dot{\vec{r}} 
\right) \cdot [\vec{H}\times \vec{r}] + \frac{e^2}{\varepsilon r},
\end{equation}
where $M = m_e + m_h$ and $\mu = \frac{m_e m_h}{m_e+m_h}$ are, respectively, the total and the reduced mass of the electron-hole pair, 
\begin{equation}
\gamma=\frac{m_h-m_e}{m_h+m_e},
\end{equation}
and the total time derivative $\frac{e}{2c} \frac{d}{dt} \left( \vec{r} \cdot [\vec{H}\times \vec{R}] \right)$ was omitted 
from the Lagrangian.

The corresponding Hamiltonian is
\begin{equation}
H =\frac{1}{2M}\left(\vec{P}+\frac{e}{c}[\vec{H}\times \vec{r}]\right)^2
+\frac{1}{2\mu}\left(\vec{p}+\frac{e\gamma}{2c}[\vec{H}\times \vec{r}]\right)^2-\frac{e^2}{\varepsilon r},
\end{equation}
where $\vec{P}=M\dot{\vec{R}}-\frac{e}{c}[\vec{H}\times\vec{r}]$ and 
$\vec{p}=\mu \dot{\vec{r}}-\frac{e\gamma}{2c}\left[\vec{H}\times\vec{r}\right]$ are, respectively, the total and relative momenta. 
The Hamiltonian is independent of the center-of-mass coordinate $\vec{R}$, which makes the total momentum $\vec{P}$ an integral 
of motion. Since only the excitons with ${\bf P} = 0$ are directly excited in an optical experiment, the Hamiltonian can be 
written in the form,
\begin{equation}
H =\frac{\vec{p}^2}{2\mu} + \frac{e}
{2\mu c} \vec{L} \cdot (\gamma \vec{H}) 
-\frac{e^2}{\varepsilon r}+ \frac{e^2}{8\mu c^2} \left[\vec{H}\times \vec{r}\right]^2,
\label{eq:hydrogen}
\end{equation}
where $\vec{L} = \left[\vec{r} \times \vec{p} \right]$ is the orbital momentum. Equation~(\ref{eq:hydrogen}) has the 
form of the Hamiltonian of an electron in the hydrogen atom in a magnetic field $\gamma\vec{H}$ and in a parabolic trapping 
potential in the plane perpendicular to $\vec{H}$ (the last term in Eq.~(\ref{eq:hydrogen}) also known as the Langevin or 
diamagnetic term).

For convenience we choose the cylindrical coordinates with the $z$ axis along the magnetic field, and $\rho = \sqrt{x^2 + y^2}$. 
The Hamiltonian (\ref{eq:hydrogen}) is invariant under rotations around the direction of magnetic field, therefore 
$m = \frac{1}{\hbar}L_z$ is a good quantum number. As was discussed in Sec.~\ref{sec:selection}, only the exciton states 
with $m = 0,\pm 1$ are excited in the photoabsorption experiment. 

The dependence of eigenfunctions on $z$ and $\rho$ was found numerically by solving eigenvalue problem for the 
Hamiltonian written in the basis of functions, 
\begin{equation}
\psi_{mn_{\rho}n_{z}}(\rho,z) = e^{-\frac{(z^2+\rho^2)}{2l^2}} \left(\frac{\rho}{l}\right)^{|m|}  
L_{n_{\rho}}^{|m|}\left(\frac{\rho^2}{l^2}\right)
H_{n_z}\left(\frac{z}{l}\right),
\label{eq:basis} 
\end{equation}
where $l = \sqrt{\frac{2 \hbar c}{e H}}$ is the magnetic length $l_0 = \sqrt{\frac{\hbar c}{e H}}$ multiplied by $\sqrt{2}$, 
while $H_{n_{z}}$ and $L_{n_{\rho}}^{|m|}$ are, respectively, the Hermite and Laguerre polynomials. In this basis the matrix 
elements of the Coulomb interaction can be evaluated analytically, which simplifies the calculation of the eigenstates of the 
Hamiltonian (\ref{eq:hydrogen}).

This method is appropriate in the strong-field limit, where the distance between the Landau levels, $\frac{\hbar e H}{\mu c}$, is 
larger than the exciton Rydberg constant, ${\rm Ry}_{X}=\frac{\mu e^4}{2\varepsilon^2 \hbar^2}$. However, using a rather large basis 
with $n_{\rho} \leq 10$ and $n_z \leq 10$, we can extend its applicability up to the physically interesting fields of $\sim 15$T. 
In the opposite limit of weak fields we  diagonalize the Hamiltonian (\ref{eq:hydrogen}) in the basis of the zero-field hydrogen wave 
functions of the discrete spectrum and truncate the basis at $n = 20$. In both cases we checked that the energies of the levels do not 
change upon a further increase of the basis dimension. The $H$-dependence of the exciton energies 
obtained in the two opposite limits matches in the region of intermediate magnetic fields, which allows us to calculate the exciton 
energies for arbitrary magnetic fields. The dashed lines in figures~\ref{fig:fit} show the magnetic field dependence of the excitonic 
levels calculated by numerical diagonalization of the Hamiltonian \mathref{eq:hydrogen} superimposed on experimental absorption spectra. 
The red dots indicate the points of a crossover between high- and low-field lines.

\section{\label{sec:fit}Fit to experimental data}
In order to fit the experimental data, it is necessary to take into account the field-dependence of exciton energies resulting from 
the interaction of the electron and hole spins with the magnetic field $H \| z$:
\begin{equation}
\hat{H}_{\rm spin}=\mu_\mathrm{B} H(g_c j_e^z+g_v j_v^z),
\end{equation}
where $j_c^z$ and $j_v^z$ are the $z$-components of the angular momenta of the conduction and valence electron forming the 
exciton and $g_c$ and $g_v$ are respectively the $g$-factors of electrons in the conduction and valence band. The interaction 
of spins with the magnetic field mixes the ortho \ket{1,0} and para \ket{0,0} states and the corresponding energies are:
$$E_\pm=E_0\pm\sqrt{\left(\frac{\Delta_{\rm o-p}}{2}\right)^2+\left[\frac{1}{2}(g_c-g_v)\mu_\mathrm{B}B\right]^2}$$
where $\Delta_{\rm o-p}$ is the exchange splitting between the ortho and para states in zero field. It is proportional to 
the square of the enveloping electron-hole wavefunction $\varphi$ at $r=0$ \cite{Kavoulakis97}, which is only nonzero for $s$-states, 
whereas electric quadrupole excitation is only allowed to $p$-states [see Eq.~(\ref{eq:selrule1})]. In fact, existing experimental data 
on the yellow exciton series shows that the ortho-para splitting is zero for $n > 1$ within the experimental precision \cite{jorger05}.

Therefore, in an applied magnetic field the spin part of exciton wave functions has the form,
\begin{eqnarray}\nonumber
\psi_+&=&\frac{\ket{10}+\ket{00}}{\sqrt{2}}=\ket{\uparrow_c}\ket{\downarrow_v}, \\ 
\psi_-&=&\frac{\ket{10}-\ket{00}}{\sqrt{2}}=\ket{\downarrow_c}\ket{\uparrow_v},
\label{opwf}
\end{eqnarray}
which allows us to extract the $g$-factors of electrons in the conduction and valence band (see Sec.~\ref{sec:discussion}).

Furthermore, to extract the exciton parameters it is important to take into account that the coupling of excitons to the 
lattice modifies the shape of the absorption peaks, and shifts the maximum of the absorption away from the position in the 
rigid lattice. The lineshape can be fitted with the asymmetric Lorenzian\cite{toy1964},
\begin{equation}\label{shape}
I(\omega)\sim\frac{\hbar\Gamma/2+2A(\hbar\omega-E)}{(\hbar\omega-E)^2+(\hbar\Gamma/2)^2},
\end{equation}
where $E$ is the exciton energy in the ``rigid'' lattice, $\Gamma$ is the exciton-phonon scattering strength and $A$ is 
the asymmetry parameter.

\begin{figure}[ht]
 \centering
 \includegraphics[width=\columnwidth]{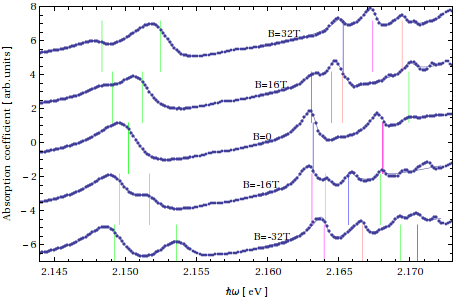}
\caption{(color online) The absorption spectrum for the $n = 2,3$ exciton (dots) at various values of magnetic field 
fitted with the sum of asymmetric Lorenzian peaks (solid line) The `bare' exciton energies (see text) are indicated by vertical lines.
}
\label{fig:fitN2}
\end{figure}

The fit of the absorption spectrum for the $n = 2,3$ excitons at various values of magnetic field is shown in Fig.~\ref{fig:fitN2}, 
where circles represent the experimental data and the continuous line is a fit by Eq.~(\ref{shape}). The linewidth $\Gamma=2$meV 
for $n=2$ levels agrees with the results of earlier studies \cite{zverev61}.

The excellent quality of the fit allows us to extract the `bare' exciton energies, indicated by vertical lines. Since the maxima 
of the absorption spectra are displaced with respect to the bare exciton energies, this procedure enables us to extract the 
$g$-factors and masses of electron and hole from experimental data in a more reliable way.

\section{\label{sec:discussion}Discussion}

\begin{figure*}[ht]
\begin{minipage}{.48\linewidth}
 \centering
 a)\includegraphics[width=0.95\columnwidth]{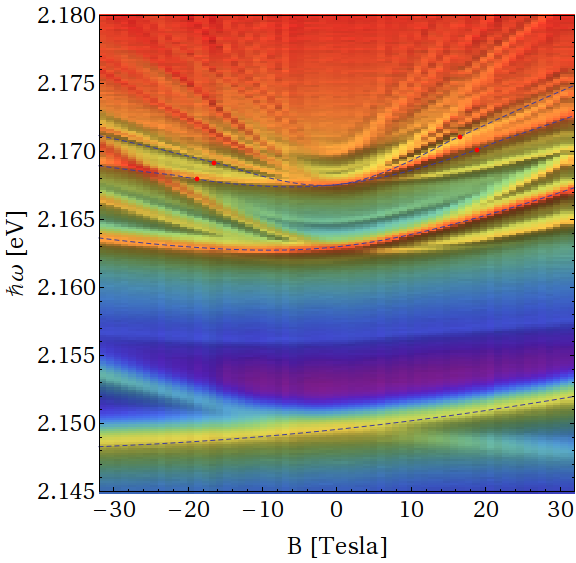}
\end{minipage}
\hspace{.02\linewidth}
\begin{minipage}{.48\linewidth}
 \centering
 b)\includegraphics[width=0.95\columnwidth]{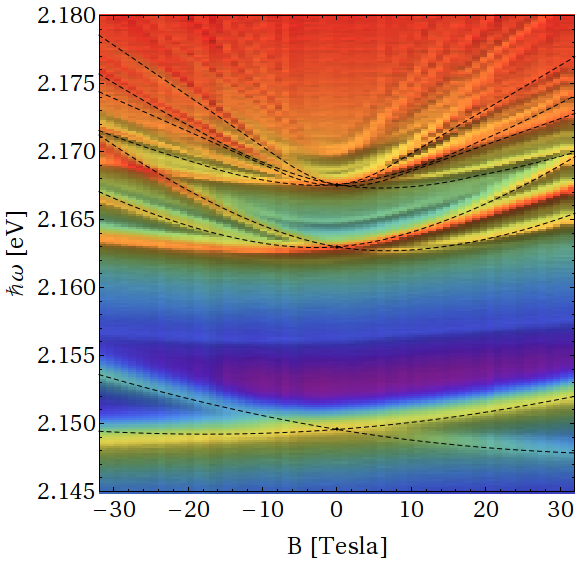}
\end{minipage}
\caption{\label{fig:fit}(color online) The magnetoabsorption spectra measured in the Faraday geometry, $\vec{H} \| \vec{q} \|(001)$, 
together with the theoretically calculated magnetic field dependence of the energies (dashed lines) for the excitons with 
[left panel a), set 1] $S_z = -1$ and $m = 0$ and [right panel b), set 2] $S_z = 0$ and $m = -1$.}
\end{figure*}

Figure~\ref{fig:fit} shows the magnetic field and photon energy dependence of the optical absorption with the calculated 
excitonic energies superimposed. In accordance with the selection rules, $n\geq 2$ excitons contribute to the optical 
absorption, forming at zero magnetic field a hydrogen-like series $\hbar\omega_n=E_{gap}-\frac{\mathrm Ry_X}{n^2}$ with 
the optical band gap $E_{gap}=2.174$~eV and the excitonic Rydberg constant ${\mathrm Ry_X}=98$~meV. \footnote{The binding 
energy of $n=1$ exciton is anomalously large ($150$~meV). The exciton radius of the $n=1$ exciton (7~\AA) is comparable to 
the lattice constant (4.2~\AA), which leads to significant central cell corrections and reduced screening of Coulomb 
interaction responsible for this anomaly \cite{Kavoulakis97}. The corrections to the binding energy of the $n=2$ level, produced 
by these mechanisms, are negligible.} Using $\epsilon=7.5$ for the dielectric constant \cite{Halpern67,Fishman09,hodby76}, we 
obtain the reduced mass $\mu=0.41 m_0$ in agreement with Ref. \cite{zhil69}.

According to the selection rules derived in Sec.~\ref{sec:selection}, the absorption spectrum for the right circularly-polarized 
light ($\lambda=+1$) is formed by two different sets of states: the states with $S_z= -1$ and $m=0$ (set 1) and the states 
with $S_z=0$ and $m= + 1$ (set 2). Dashed lines in Fig.~\ref{fig:fit} show the numerically calculated energies of $n=2,3,4$ 
exciton states in magnetic field up to $32$T, which belong, respectively, to the sets 1 and 2, superimposed on the experimental absorption spectra.

Set 1 corresponds to the absorption of a photon with $\lambda=+1$ and creation of an exciton in the state $\ket{1,-1},m=0$. 
The magnetic moment in this state is determined by the atomic g-factors of electrons and holes. Since the hole in the upper 
valence band has $s_z=1/2$ and $l_z=-1$, it has zero g-factor since $(l_z+2s_z)=0$ \cite{Ballhausen}. The electron wave function 
is mostly of Cu $s$ character, and since in this case the spin-orbit interaction is not effective, the g-factor should be close to 
the bare value of $2$. Indeed, a good agreement with the experiment is obtained for $g_c=2.0$ (see Fig.~\ref{fig:fit}b).

The last term in the Hamiltonian Eq.~(\ref{eq:hydrogen}) mixes the state $\ket{l,m}$ with the states $\ket{l,m}$ and $\ket{l\pm 2,m}$. 
This leads to the mixing of $p$ and $f$ states for $n\geq 4$ giving rise to additional lines. In general the line with the main 
quantum number $n$ splits in magnetic field into $\left[\frac{n}{2}\right]$ levels (here $[x]$ denotes the largest integer smaller than $x$).

Set 2 of the absorption lines is produced by the ${\ket{1,0}\pm\ket{0,0},m=1}$ excitonic transitions. This set has twice more states, 
corresponding to $\ket{\uparrow_c\downarrow_v}$ and $\ket{\downarrow_c\uparrow_v}$. 
The energy shifts of these levels up to the terms linear in the magnetic field are
$$E=2\left(\frac{m_0}{m_h}-\frac{m_0}{m_e}\right)\mu_B H m\pm\frac{1}{2}|g_c-g_v|\mu_B H.$$
We extracted $m_e=1.0m_0, m_h=0.7m_0, |g_c-g_v|=2.25$ and $g_c+g_v=2.0$, so that for the atomic g-factors of electrons in the 
conduction and valence bands we obtain, respectively, $g_c=2.1, g_v=-0.1$ in good agreement with our simple arguments given above.
The effective masses coincide with the results of the cyclotron resonance experiments \cite{hodby76}.
These values of the parameters result in good agreement between the calculated and the measured spectra.

To conclude, we studied the magneto-absorption spectrum of cuprous oxide in high magnetic fields, calculated excitonic energies 
for arbitrary field values, and extracted the exciton parameters from the intermediate field region, where the peaks are clearly 
discernible. Our results suggest that the wide $n=2$ line is a result of the overlap of two lines with different quantum numbers, 
resolving a long-standing controversy over the number of excitonic levels in the applied magnetic field. This observation allows us 
to extract the masses of electrons and holes, which are consistent with the results of cyclotron resonance experiments, and g-factors 
consistent with the present understanding of the nature of valence and conduction bands of Cu$_2$O.

This work was partially supported by the European Access Program RITA-CT-2003-505474 and Grenoble High Magnetic Fields Laboratory (Grenoble, France).


\begin{thebibliography}{30}%
\makeatletter
\providecommand \@ifxundefined [1]{%
 \@ifx{#1\undefined}
}%
\providecommand \@ifnum [1]{%
 \ifnum #1\expandafter \@firstoftwo
 \else \expandafter \@secondoftwo
 \fi
}%
\providecommand \@ifx [1]{%
 \ifx #1\expandafter \@firstoftwo
 \else \expandafter \@secondoftwo
 \fi
}%
\providecommand \natexlab [1]{#1}%
\providecommand \enquote  [1]{``#1''}%
\providecommand \bibnamefont  [1]{#1}%
\providecommand \bibfnamefont [1]{#1}%
\providecommand \citenamefont [1]{#1}%
\providecommand \href@noop [0]{\@secondoftwo}%
\providecommand \href [0]{\begingroup \@sanitize@url \@href}%
\providecommand \@href[1]{\@@startlink{#1}\@@href}%
\providecommand \@@href[1]{\endgroup#1\@@endlink}%
\providecommand \@sanitize@url [0]{\catcode `\\12\catcode `\$12\catcode
  `\&12\catcode `\#12\catcode `\^12\catcode `\_12\catcode `\%12\relax}%
\providecommand \@@startlink[1]{}%
\providecommand \@@endlink[0]{}%
\providecommand \url  [0]{\begingroup\@sanitize@url \@url }%
\providecommand \@url [1]{\endgroup\@href {#1}{\urlprefix }}%
\providecommand \urlprefix  [0]{URL }%
\providecommand \Eprint [0]{\href }%
\providecommand \doibase [0]{http://dx.doi.org/}%
\providecommand \selectlanguage [0]{\@gobble}%
\providecommand \bibinfo  [0]{\@secondoftwo}%
\providecommand \bibfield  [0]{\@secondoftwo}%
\providecommand \translation [1]{[#1]}%
\providecommand \BibitemOpen [0]{}%
\providecommand \bibitemStop [0]{}%
\providecommand \bibitemNoStop [0]{.\EOS\space}%
\providecommand \EOS [0]{\spacefactor3000\relax}%
\providecommand \BibitemShut  [1]{\csname bibitem#1\endcsname}%
\let\auto@bib@innerbib\@empty
\bibitem [{\citenamefont {Wannier}(1937)}]{Wannier37}%
  \BibitemOpen
  \bibfield  {author} {\bibinfo {author} {\bibfnamefont {G.~H.}\ \bibnamefont
  {Wannier}},\ }\href {\doibase 10.1103/PhysRev.52.191} {\bibfield  {journal}
  {\bibinfo  {journal} {Phys. Rev.}\ }\textbf {\bibinfo {volume} {52}},\
  \bibinfo {pages} {191} (\bibinfo {year} {1937})}\BibitemShut {NoStop}%
\bibitem [{\citenamefont {Elliott}(1961)}]{Elliott61}%
  \BibitemOpen
  \bibfield  {author} {\bibinfo {author} {\bibfnamefont {R.}~\bibnamefont
  {Elliott}},\ }\href {\doibase 10.1103/PhysRev.124.340} {\bibfield  {journal}
  {\bibinfo  {journal} {Phys. Rev.}\ }\textbf {\bibinfo {volume} {124}},\
  \bibinfo {pages} {340} (\bibinfo {year} {1961})}\BibitemShut {NoStop}%
\bibitem [{\citenamefont {Kavoulakis}\ \emph {et~al.}(1997)\citenamefont
  {Kavoulakis}, \citenamefont {Chang},\ and\ \citenamefont
  {Baym}}]{Kavoulakis97}%
  \BibitemOpen
  \bibfield  {author} {\bibinfo {author} {\bibfnamefont {G.~M.}\ \bibnamefont
  {Kavoulakis}}, \bibinfo {author} {\bibfnamefont {Y.-C.}\ \bibnamefont
  {Chang}}, \ and\ \bibinfo {author} {\bibfnamefont {G.}~\bibnamefont {Baym}},\
  }\href {\doibase 10.1103/PhysRevB.55.7593} {\bibfield  {journal} {\bibinfo
  {journal} {Phys. Rev. B}\ }\textbf {\bibinfo {volume} {55}},\ \bibinfo
  {pages} {7593} (\bibinfo {year} {1997})}\BibitemShut {NoStop}%
\bibitem [{\citenamefont {Keldysh}\ and\ \citenamefont
  {Kozlov}(1968)}]{Keldysh68}%
  \BibitemOpen
  \bibfield  {author} {\bibinfo {author} {\bibfnamefont {L.}~\bibnamefont
  {Keldysh}}\ and\ \bibinfo {author} {\bibfnamefont {A.}~\bibnamefont
  {Kozlov}},\ }\href@noop {} {\bibfield  {journal} {\bibinfo  {journal} {JETP}\
  }\textbf {\bibinfo {volume} {27}},\ \bibinfo {pages} {521} (\bibinfo {year}
  {1968})}\BibitemShut {NoStop}%
\bibitem [{\citenamefont {Yoshioka}\ \emph {et~al.}(2011)\citenamefont
  {Yoshioka}, \citenamefont {Chae},\ and\ \citenamefont
  {Kuwata-Gonokami}}]{Kuwata11}%
  \BibitemOpen
  \bibfield  {author} {\bibinfo {author} {\bibfnamefont {K.}~\bibnamefont
  {Yoshioka}}, \bibinfo {author} {\bibfnamefont {E.}~\bibnamefont {Chae}}, \
  and\ \bibinfo {author} {\bibfnamefont {M.}~\bibnamefont {Kuwata-Gonokami}},\
  }\href@noop {} {\bibfield  {journal} {\bibinfo  {journal} {Nature Commun.}\
  }\textbf {\bibinfo {volume} {2}},\ \bibinfo {pages} {328} (\bibinfo {year}
  {2011})}\BibitemShut {NoStop}%
\bibitem [{\citenamefont {Snoke}(2002)}]{Snoke02}%
  \BibitemOpen
  \bibfield  {author} {\bibinfo {author} {\bibfnamefont {D.}~\bibnamefont
  {Snoke}},\ }\href@noop {} {\bibfield  {journal} {\bibinfo  {journal}
  {Science}\ }\textbf {\bibinfo {volume} {298}},\ \bibinfo {pages} {1368}
  (\bibinfo {year} {2002})}\BibitemShut {NoStop}%
\bibitem [{\citenamefont {Griffin}\ \emph {et~al.}(1995)\citenamefont
  {Griffin}, \citenamefont {Snoke},\ and\ \citenamefont {Stringari}}]{Snoke95}%
  \BibitemOpen
  \bibfield  {author} {\bibinfo {author} {\bibfnamefont {A.}~\bibnamefont
  {Griffin}}, \bibinfo {author} {\bibfnamefont {D.~W.}\ \bibnamefont {Snoke}},
  \ and\ \bibinfo {author} {\bibfnamefont {S.}~\bibnamefont {Stringari}},\
  }\href@noop {} {\emph {\bibinfo {title} {Bose-Einstein Condensation}}}\
  (\bibinfo  {publisher} {Cambridge University Press},\ \bibinfo {address}
  {Cambridge},\ \bibinfo {year} {1995})\BibitemShut {NoStop}%
\bibitem [{\citenamefont {Halpern}\ and\ \citenamefont
  {Zakharchenya}(1967)}]{Halpern67}%
  \BibitemOpen
  \bibfield  {author} {\bibinfo {author} {\bibfnamefont {J.}~\bibnamefont
  {Halpern}}\ and\ \bibinfo {author} {\bibfnamefont {B.~P.}\ \bibnamefont
  {Zakharchenya}},\ }\href {\doibase DOI: 10.1016/0038-1098(67)90081-6}
  {\bibfield  {journal} {\bibinfo  {journal} {Solid State Commun.}\ }\textbf
  {\bibinfo {volume} {5}},\ \bibinfo {pages} {633} (\bibinfo {year}
  {1967})}\BibitemShut {NoStop}%
\bibitem [{\citenamefont {Zhilich}\ \emph {et~al.}(1969)\citenamefont
  {Zhilich}, \citenamefont {Halpern},\ and\ \citenamefont
  {Zakharchenya}}]{zhil69}%
  \BibitemOpen
  \bibfield  {author} {\bibinfo {author} {\bibfnamefont {A.~G.}\ \bibnamefont
  {Zhilich}}, \bibinfo {author} {\bibfnamefont {J.}~\bibnamefont {Halpern}}, \
  and\ \bibinfo {author} {\bibfnamefont {B.~P.}\ \bibnamefont {Zakharchenya}},\
  }\href {\doibase 10.1103/PhysRev.188.1294} {\bibfield  {journal} {\bibinfo
  {journal} {Phys. Rev.}\ }\textbf {\bibinfo {volume} {188}},\ \bibinfo {pages}
  {1294} (\bibinfo {year} {1969})}\BibitemShut {NoStop}%
\bibitem [{\citenamefont {Hodby}\ \emph {et~al.}(1976)\citenamefont {Hodby},
  \citenamefont {Jenkins}, \citenamefont {Schwab}, \citenamefont {Tamura},\
  and\ \citenamefont {Trivich}}]{hodby76}%
  \BibitemOpen
  \bibfield  {author} {\bibinfo {author} {\bibfnamefont {J.~W.}\ \bibnamefont
  {Hodby}}, \bibinfo {author} {\bibfnamefont {T.~E.}\ \bibnamefont {Jenkins}},
  \bibinfo {author} {\bibfnamefont {C.}~\bibnamefont {Schwab}}, \bibinfo
  {author} {\bibfnamefont {H.}~\bibnamefont {Tamura}}, \ and\ \bibinfo {author}
  {\bibfnamefont {D.}~\bibnamefont {Trivich}},\ }\href@noop {} {\bibfield
  {journal} {\bibinfo  {journal} {J. Phys. C}\ }\textbf {\bibinfo {volume}
  {9}},\ \bibinfo {pages} {1429} (\bibinfo {year} {1976})}\BibitemShut
  {NoStop}%
\bibitem [{\citenamefont {Sasaki}\ and\ \citenamefont
  {Kuwabara}(1973)}]{sas73}%
  \BibitemOpen
  \bibfield  {author} {\bibinfo {author} {\bibfnamefont {H.}~\bibnamefont
  {Sasaki}}\ and\ \bibinfo {author} {\bibfnamefont {G.}~\bibnamefont
  {Kuwabara}},\ }\href {\doibase 10.1143/JPSJ.34.95} {\bibfield  {journal}
  {\bibinfo  {journal} {J. Phys. Soc. Jpn.}\ }\textbf {\bibinfo {volume}
  {34}},\ \bibinfo {pages} {95} (\bibinfo {year} {1973})}\BibitemShut {NoStop}%
\bibitem [{\citenamefont {Kobayashi}\ \emph {et~al.}(1989)\citenamefont
  {Kobayashi}, \citenamefont {Kanisawa}, \citenamefont {Misu}, \citenamefont
  {Nagamune}, \citenamefont {Takeyama},\ and\ \citenamefont {Miura}}]{kob89}%
  \BibitemOpen
  \bibfield  {author} {\bibinfo {author} {\bibfnamefont {M.}~\bibnamefont
  {Kobayashi}}, \bibinfo {author} {\bibfnamefont {K.}~\bibnamefont {Kanisawa}},
  \bibinfo {author} {\bibfnamefont {A.}~\bibnamefont {Misu}}, \bibinfo {author}
  {\bibfnamefont {Y.}~\bibnamefont {Nagamune}}, \bibinfo {author}
  {\bibfnamefont {S.}~\bibnamefont {Takeyama}}, \ and\ \bibinfo {author}
  {\bibfnamefont {N.}~\bibnamefont {Miura}},\ }\href@noop {} {\bibfield
  {journal} {\bibinfo  {journal} {J. Phys. Soc. Jpn.}\ }\textbf {\bibinfo
  {volume} {58}},\ \bibinfo {pages} {1823} (\bibinfo {year}
  {1989})}\BibitemShut {NoStop}%
\bibitem [{\citenamefont {Seyama}\ \emph {et~al.}(2003)\citenamefont {Seyama},
  \citenamefont {Takamasu}, \citenamefont {Imanaka}, \citenamefont {Yamaguchi},
  \citenamefont {Masumi},\ and\ \citenamefont {Kido}}]{sey03}%
  \BibitemOpen
  \bibfield  {author} {\bibinfo {author} {\bibfnamefont {M.}~\bibnamefont
  {Seyama}}, \bibinfo {author} {\bibfnamefont {T.}~\bibnamefont {Takamasu}},
  \bibinfo {author} {\bibfnamefont {Y.}~\bibnamefont {Imanaka}}, \bibinfo
  {author} {\bibfnamefont {H.}~\bibnamefont {Yamaguchi}}, \bibinfo {author}
  {\bibfnamefont {T.}~\bibnamefont {Masumi}}, \ and\ \bibinfo {author}
  {\bibfnamefont {G.}~\bibnamefont {Kido}},\ }\href@noop {} {\bibfield
  {journal} {\bibinfo  {journal} {J. Phys. Soc. Jpn.}\ }\textbf {\bibinfo
  {volume} {72}},\ \bibinfo {pages} {437} (\bibinfo {year} {2003})}\BibitemShut
  {NoStop}%
\bibitem [{\citenamefont {Gross}\ and\ \citenamefont
  {Zakharchenya}(1956)}]{gross56}%
  \BibitemOpen
  \bibfield  {author} {\bibinfo {author} {\bibfnamefont {E.~F.}\ \bibnamefont
  {Gross}}\ and\ \bibinfo {author} {\bibfnamefont {B.~P.}\ \bibnamefont
  {Zakharchenya}},\ }\href@noop {} {\bibfield  {journal} {\bibinfo  {journal}
  {Dokl. Akad. Nauk SSSR[Sov. Phys.--- Doklady]}\ }\textbf {\bibinfo {volume}
  {111}},\ \bibinfo {pages} {564} (\bibinfo {year} {1956})}\BibitemShut
  {NoStop}%
\bibitem [{\citenamefont {Hammura}\ \emph {et~al.}(2000)\citenamefont
  {Hammura}, \citenamefont {Sakai},\ and\ \citenamefont {Seyama}}]{ham00}%
  \BibitemOpen
  \bibfield  {author} {\bibinfo {author} {\bibfnamefont {K.}~\bibnamefont
  {Hammura}}, \bibinfo {author} {\bibfnamefont {K.}~\bibnamefont {Sakai}}, \
  and\ \bibinfo {author} {\bibfnamefont {M.}~\bibnamefont {Seyama}},\
  }\href@noop {} {\bibfield  {journal} {\bibinfo  {journal} {Prog. Theor. Phys.
  Suppl.}\ }\textbf {\bibinfo {volume} {138}},\ \bibinfo {pages} {143}
  (\bibinfo {year} {2000})}\BibitemShut {NoStop}%
\bibitem [{\citenamefont {Naka}\ \emph {et~al.}(2012)\citenamefont {Naka},
  \citenamefont {Akimoto}, \citenamefont {Shirai},\ and\ \citenamefont
  {Kan'no}}]{Naka12}%
  \BibitemOpen
  \bibfield  {author} {\bibinfo {author} {\bibfnamefont {N.}~\bibnamefont
  {Naka}}, \bibinfo {author} {\bibfnamefont {I.}~\bibnamefont {Akimoto}},
  \bibinfo {author} {\bibfnamefont {M.}~\bibnamefont {Shirai}}, \ and\ \bibinfo
  {author} {\bibfnamefont {K.-i.}\ \bibnamefont {Kan'no}},\ }\href {\doibase
  10.1103/PhysRevB.85.035209} {\bibfield  {journal} {\bibinfo  {journal} {Phys.
  Rev. B}\ }\textbf {\bibinfo {volume} {85}},\ \bibinfo {pages} {035209}
  (\bibinfo {year} {2012})}\BibitemShut {NoStop}%
\bibitem [{\citenamefont {Schmidt-Withley}\ \emph {et~al.}(1974)\citenamefont
  {Schmidt-Withley}, \citenamefont {Martinez-Clemente},\ and\ \citenamefont
  {Revcolevschi}}]{sch74}%
  \BibitemOpen
  \bibfield  {author} {\bibinfo {author} {\bibfnamefont {R.~D.}\ \bibnamefont
  {Schmidt-Withley}}, \bibinfo {author} {\bibfnamefont {M.}~\bibnamefont
  {Martinez-Clemente}}, \ and\ \bibinfo {author} {\bibfnamefont
  {A.}~\bibnamefont {Revcolevschi}},\ }\href@noop {} {\bibfield  {journal}
  {\bibinfo  {journal} {J. Cryst. Growth}\ }\textbf {\bibinfo {volume} {23}},\
  \bibinfo {pages} {113} (\bibinfo {year} {1974})}\BibitemShut {NoStop}%
\bibitem [{\citenamefont {Uihlein}\ \emph {et~al.}(1981)\citenamefont
  {Uihlein}, \citenamefont {Fr{\"o}hlich},\ and\ \citenamefont
  {Kenklies}}]{Uihlein81}%
  \BibitemOpen
  \bibfield  {author} {\bibinfo {author} {\bibfnamefont {C.}~\bibnamefont
  {Uihlein}}, \bibinfo {author} {\bibfnamefont {D.}~\bibnamefont
  {Fr{\"o}hlich}}, \ and\ \bibinfo {author} {\bibfnamefont {R.}~\bibnamefont
  {Kenklies}},\ }\href {\doibase 10.1103/PhysRevB.23.2731} {\bibfield
  {journal} {\bibinfo  {journal} {Phys. Rev. B}\ }\textbf {\bibinfo {volume}
  {23}},\ \bibinfo {pages} {2731} (\bibinfo {year} {1981})}\BibitemShut
  {NoStop}%
\bibitem [{\citenamefont {Kiselev}\ and\ \citenamefont
  {Zhilich}(1971)}]{kiselev71}%
  \BibitemOpen
  \bibfield  {author} {\bibinfo {author} {\bibfnamefont {V.~A.}\ \bibnamefont
  {Kiselev}}\ and\ \bibinfo {author} {\bibfnamefont {A.~G.}\ \bibnamefont
  {Zhilich}},\ }\href@noop {} {\bibfield  {journal} {\bibinfo  {journal} {Fiz.
  Tverd. Tela}\ }\textbf {\bibinfo {volume} {13}},\ \bibinfo {pages} {2398}
  (\bibinfo {year} {1971})}\BibitemShut {NoStop}%
\bibitem [{\citenamefont {Pikus}\ and\ \citenamefont {Bir}(1971)}]{pikus71}%
  \BibitemOpen
  \bibfield  {author} {\bibinfo {author} {\bibfnamefont {G.}~\bibnamefont
  {Pikus}}\ and\ \bibinfo {author} {\bibfnamefont {G.}~\bibnamefont {Bir}},\
  }\href@noop {} {\bibfield  {journal} {\bibinfo  {journal} {Zh. Eksp. Teor.
  Fiz.}\ }\textbf {\bibinfo {volume} {60}},\ \bibinfo {pages} {195} (\bibinfo
  {year} {1971})}\BibitemShut {NoStop}%
\bibitem [{\citenamefont {Denisov}\ and\ \citenamefont
  {Makarov}(1973)}]{denisov73}%
  \BibitemOpen
  \bibfield  {author} {\bibinfo {author} {\bibfnamefont {M.}~\bibnamefont
  {Denisov}}\ and\ \bibinfo {author} {\bibfnamefont {V.}~\bibnamefont
  {Makarov}},\ }\href@noop {} {\bibfield  {journal} {\bibinfo  {journal} {Phys.
  Status Solidi B}\ }\textbf {\bibinfo {volume} {56}},\ \bibinfo {pages} {9}
  (\bibinfo {year} {1973})}\BibitemShut {NoStop}%
\bibitem [{\citenamefont {Fishman}\ \emph {et~al.}(2009)\citenamefont
  {Fishman}, \citenamefont {Faugeras}, \citenamefont {Potemski}, \citenamefont
  {Revcolevschi},\ and\ \citenamefont {{van Loosdrecht}}}]{Fishman09}%
  \BibitemOpen
  \bibfield  {author} {\bibinfo {author} {\bibfnamefont {D.}~\bibnamefont
  {Fishman}}, \bibinfo {author} {\bibfnamefont {C.}~\bibnamefont {Faugeras}},
  \bibinfo {author} {\bibfnamefont {M.}~\bibnamefont {Potemski}}, \bibinfo
  {author} {\bibfnamefont {A.}~\bibnamefont {Revcolevschi}}, \ and\ \bibinfo
  {author} {\bibfnamefont {P.~H.~M.}\ \bibnamefont {{van Loosdrecht}}},\ }\href
  {\doibase 10.1103/PhysRevB.80.045208} {\bibfield  {journal} {\bibinfo
  {journal} {Phys. Rev. B}\ }\textbf {\bibinfo {volume} {80}},\ \bibinfo {eid}
  {045208} (\bibinfo {year} {2009})}\BibitemShut {NoStop}%
\bibitem [{\citenamefont {Gor'kov}\ and\ \citenamefont
  {Dzyaloshinskii}(1967)}]{GorDzya1968}%
  \BibitemOpen
  \bibfield  {author} {\bibinfo {author} {\bibfnamefont {L.}~\bibnamefont
  {Gor'kov}}\ and\ \bibinfo {author} {\bibfnamefont {I.}~\bibnamefont
  {Dzyaloshinskii}},\ }\href@noop {} {\bibfield  {journal} {\bibinfo  {journal}
  {JETP}\ }\textbf {\bibinfo {volume} {26}},\ \bibinfo {pages} {449} (\bibinfo
  {year} {1967})}\BibitemShut {NoStop}%
\bibitem [{\citenamefont {Lozovik}\ and\ \citenamefont
  {Volkov}(2003)}]{lozovik03}%
  \BibitemOpen
  \bibfield  {author} {\bibinfo {author} {\bibfnamefont {Y.}~\bibnamefont
  {Lozovik}}\ and\ \bibinfo {author} {\bibfnamefont {S.}~\bibnamefont
  {Volkov}},\ }\href@noop {} {\bibfield  {journal} {\bibinfo  {journal} {JETP}\
  }\textbf {\bibinfo {volume} {96}},\ \bibinfo {pages} {564} (\bibinfo {year}
  {2003})}\BibitemShut {NoStop}%
\bibitem [{Note1()}]{Note1}%
  \BibitemOpen
  \bibinfo {note} {For a review see I.B. Khriplovich, G.Yu. Ruban,
  arXiv:quant-ph/0309014v2}\BibitemShut {NoStop}%
\bibitem [{\citenamefont {J{\"o}rger}\ \emph {et~al.}(2005)\citenamefont
  {J{\"o}rger}, \citenamefont {Fleck}, \citenamefont {Klingshirn},\ and\
  \citenamefont {{von Baltz}}}]{jorger05}%
  \BibitemOpen
  \bibfield  {author} {\bibinfo {author} {\bibfnamefont {M.}~\bibnamefont
  {J{\"o}rger}}, \bibinfo {author} {\bibfnamefont {T.}~\bibnamefont {Fleck}},
  \bibinfo {author} {\bibfnamefont {C.}~\bibnamefont {Klingshirn}}, \ and\
  \bibinfo {author} {\bibfnamefont {R.}~\bibnamefont {{von Baltz}}},\ }\href
  {\doibase 10.1103/PhysRevB.71.235210} {\bibfield  {journal} {\bibinfo
  {journal} {Phys. Rev. B}\ }\textbf {\bibinfo {volume} {71}},\ \bibinfo
  {pages} {235210} (\bibinfo {year} {2005})}\BibitemShut {NoStop}%
\bibitem [{\citenamefont {Toyozawa}(1964)}]{toy1964}%
  \BibitemOpen
  \bibfield  {author} {\bibinfo {author} {\bibfnamefont {Y.}~\bibnamefont
  {Toyozawa}},\ }\href@noop {} {\bibfield  {journal} {\bibinfo  {journal} {J.
  Phys. Chem. Solids}\ }\textbf {\bibinfo {volume} {25}},\ \bibinfo {pages}
  {59} (\bibinfo {year} {1964})}\BibitemShut {NoStop}%
\bibitem [{\citenamefont {Zverev}\ \emph {et~al.}(1961)\citenamefont {Zverev},
  \citenamefont {Noskov},\ and\ \citenamefont {Shur}}]{zverev61}%
  \BibitemOpen
  \bibfield  {author} {\bibinfo {author} {\bibfnamefont {L.~P.}\ \bibnamefont
  {Zverev}}, \bibinfo {author} {\bibfnamefont {M.~M.}\ \bibnamefont {Noskov}},
  \ and\ \bibinfo {author} {\bibfnamefont {M.~Y.}\ \bibnamefont {Shur}},\
  }\href@noop {} {\bibfield  {journal} {\bibinfo  {journal} {Sov. Phys.---
  Solid State}\ }\textbf {\bibinfo {volume} {2}},\ \bibinfo {pages} {2357}
  (\bibinfo {year} {1961})}\BibitemShut {NoStop}%
\bibitem [{Note2()}]{Note2}%
  \BibitemOpen
  \bibinfo {note} {The binding energy of $n=1$ exciton is anomalously large
  ($150$~meV). The exciton radius of the $n=1$ exciton (7~\r A) is comparable
  to the lattice constant (4.2~\r A), which leads to significant central cell
  corrections and reduced screening of Coulomb interaction responsible for this
  anomaly \cite {Kavoulakis97}. The corrections to the binding energy of the
  $n=2$ level, produced by these mechanisms, are negligible.}\BibitemShut
  {Stop}%
\bibitem [{\citenamefont {Ballhausen}(1962)}]{Ballhausen}%
  \BibitemOpen
  \bibfield  {author} {\bibinfo {author} {\bibfnamefont {C.}~\bibnamefont
  {Ballhausen}},\ }\href@noop {} {\emph {\bibinfo {title} {Introduction to
  ligand field theory}}},\ 4-i. Equivalence of $t\_{2g}$ and $e\_g$ electrons\
  (\bibinfo  {publisher} {McGRAW-HILL Book Company, INC},\ \bibinfo {year}
  {1962})\BibitemShut {NoStop}%
\end{thebibliography}
%

\end{document}